\documentclass[prl,aps,twocolumn,floatfix,floats,showpacs,fleqn]{revtex4}
\usepackage{epsfig}
\usepackage{amsmath}
\usepackage{amssymb}
\usepackage{mathrsfs}
\usepackage{color}
\newcommand{\enote}[1]{{\color{red}#1}}

\begin{document}
\title{Collective Dynamics of Dividing Chemotactic Cells}
\author{Anatolij Gelimson}
\affiliation{Rudolf Peierls Centre for Theoretical Physics, University of Oxford, Oxford OX1 3NP, UK}

\author{Ramin Golestanian}
\affiliation{Rudolf Peierls Centre for Theoretical Physics, University of Oxford, Oxford OX1 3NP, UK}
\date{\today}

\begin{abstract}
The large scale behaviour of a population of cells that grow and interact through the concentration field of the chemicals they secrete is studied using dynamical renormalization group methods. The combination of the effective long-range chemotactic interaction and lack of number conservation leads to a rich variety of phase behaviour in the system, which includes a sharp transition from a phase that has moderate (or controlled) growth and regulated chemical interactions to a phase with strong (or uncontrolled) growth and no chemical interactions. The transition point has nontrivial critical exponents. Our results might help shed light on the interplay between chemical signalling and growth in tissues and colonies, and in particular on the challenging problem of cancer metastasis.
\end{abstract}


\maketitle

Chemotactic cell motility has attracted a lot of interest in biology and medical research, as it plays an essential role in cancer metastasis \cite{Hanahan:2011}, leukocyte extravasation, angiogenesis, wound healing and embryogenesis \cite{Singer:1986}, through signalling that involves various molecules (e.g. growth factors) and is mediated by the extracellular matrix \cite{Roussos:2011}. Bacteria such as {\em E. coli} \cite{berg1} have developed an efficient run-and-tumble search strategy for the needed chemicals \cite{search} by coupling sensing of the chemicals---that is accentuated through an elaborate clustering mechanism for the protein receptors \cite{Bray}---to the motility machinery via signalling pathways that have a feedback control on the preferred direction of the rotation of the flagellar motors \cite{biochemical-circuitry-1,biochemical-circuitry-2}. The situation is more complex with eukaryotic chemotaxis, where the motility mechanisms are typically much more elaborate, e.g. involving actin polymerization or coordinated motion of thousands of molecular motors \cite{Herb}. On a more coarse-grained level, however, the resulting motion can be phenomenologically modelled as a directed mobility towards (away from) increasing concentrations of molecules that act as chemo-attractant (-repellant) \cite{KS1970}. This level of description has been successfully used to study a variety of interesting effects in bacterial behaviour such as auto-chemotaxis, where single bacteria are influenced by their own chemotactic field \cite{Tsori:2004, Grima:2005, Sengupta:2009}, and collective behaviour of bacteria caused by the chemical interactions, such as the chemotactic collapse \cite{KS1970, Chavanis:2002, Chavanis:2004} and other forms of nonequilibrium pattern formation \cite{Brenner1995,Levine2000}. Similar behaviours have been discussed for active colloids that communicate via the same type of long-range interactions \cite{Golestanian:2012,Soto:2014,Cohen:2014,Saha:2014}. Coarse-grained theories for active systems have been shown to apply to a whole variety of---seemingly unrelated---collective phenomena in biology such as flocks of birds, schools of fish, aggregations of molecular motors, and dynamic reorganization of growing tissues \cite{MCMetal2013,Vicsek1995,TonerTu1995,Karsten:2001,Chate2004,Karsten:2012,Cates:2010,Basan:2011,Montel:2011,Toner:2012,Chen:2013}.

\begin{figure} [b]
\includegraphics[width = 0.9 \columnwidth]{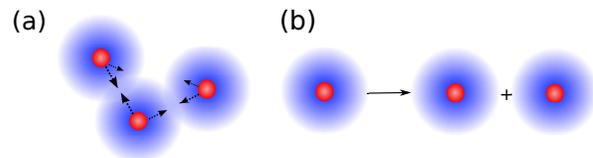}
\caption{Schematics of the model, showing (a) the interaction between the cells via a long-range field of emitted chemicals, and (b) the cell division (and death) process.}
\label{fig:spheres}
\end{figure}

One of the characteristic features of the long-time dynamics of living cells is that number conservation does not hold due to cell division and death processes, which has consequences on their collective behaviour \cite{Toner:2012}. The combined effect of this nonequilibrium property of a colony of living cells---that we model using a generic growth rule \cite{NowakBook}---and long-range chemotactic interactions among the cells is what we aim to study in this Letter. The schematics of the model is shown in Fig.~\ref{fig:spheres}. We show that in the relevant continuum description for long-time and large-scale behaviour of such a colony of cells, which we study using dynamical renormalization group (RG) methods, the two nonlinear terms representing the cell division process and chemotaxis appear at the same degree of relevance. The competition between them leads to a sharp transition from a phase that is controlled by a weakly coupled perturbatively accessible fixed point to a phase controlled by a nonaccessible strong coupling fixed point. The weakly coupled fixed point has well defined values for the strength of the chemical interaction and the growth rate. At the strong coupling fixed point, the chemical interaction becomes much less significant in competition with growth, which will collectively exhibit much larger effective rates. The weakly coupled fixed point itself corresponds to a modified chemotactic collapse transition: when the strength of the chemotactic attraction is larger than a threshold that depends on the growth rate, the cells are strongly attracted towards each other and collapse into a dense structure, while for smaller values of the chemotactic coupling the cells are dispersed into a dilute solution since the chemical attraction is not enough to overcome the diffusion. At the perturbatively accessible fixed-point, we are able to calculate critical exponents that describe a continuous phase transition. We find that the cells exhibit superdiffusive motion at the dynamical critical point, where the mean-square displacement of the cells behaves as $t^{\alpha}$, where e.g. $\alpha=1.72$ in two dimensions.

\begin{figure}[t]
\includegraphics[width =  0.62 \columnwidth]{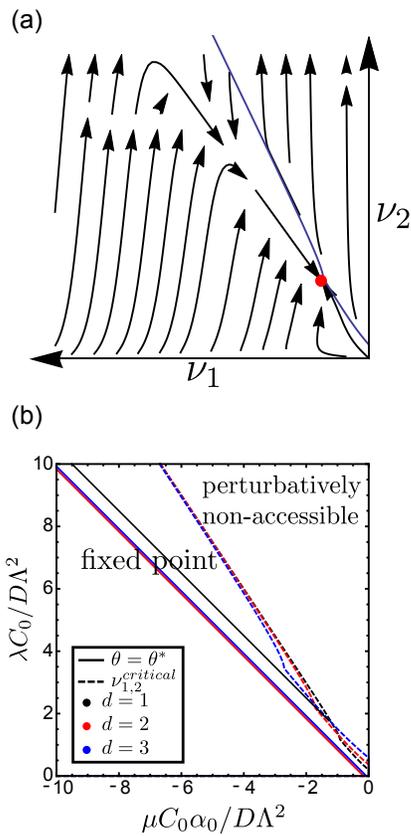}
\caption{Properties of the fixed point and the accompanying dynamical phase transition. (a) The flow around the fixed point in $(\nu_1, \nu_2)$ space. (b) The two parameter regions for $\nu_1$, $\nu_2$ for different dimensions. Above the separatrix (dashed line), a perturbatively not accessible fixed point will control the flow, whereas below it the flow will converge to the stable fixed point. The solid line represents the critical point that corresponds to the phase transition, and can be experimentally approach by tuning any of the parameters involved, e.g. the growth rate or the rate of release of the chemicals.}
\label{fig:summary2}
\end{figure}

We consider single cells that release chemicals and thus create a long-range concentration field $\phi(\mathbf{r},t)$ around them. Other cells in the suspension will then swim towards or away from the mobile chemical sources, depending on the type of cells and chemicals. Here we will assume that the response of the cells is linear with respect to the concentration gradient, such that a concentration field effectively acts as an interaction potential that leads to an effective drift. In a dissipative environment where inertial effects are negligible, the equation of motion for a single cell $i$ then reads $\partial_t \mathbf{r}_i = - \mu \left. \nabla \phi \right|_{\mathbf{r}=\mathbf{r}_i(t)}$ plus Brownian noise. Here, $\mu$ is the mobility of a cell that can be positive or negative, depending on whether the cells repel or attract each other (unit of $\mu$: $[\text{length}]^{2+d}/[\text{time}]$ in $d$ dimensions). $\phi$ obeys the diffusion equation with a source term given by the density of the cells $C(\mathbf{r}, t) = \sum_i \delta \left(\mathbf{r}-\mathbf{r}_i(t) \right)$. Since we are interested in the long-time behaviour of the system, we can assume that $\phi$ rapidly adapts to changes in $C(\mathbf{r}, t)$. In this case the potential $\phi$ is Coulomb-like, namely, $-\nabla^2 \phi = \alpha_0 C(\mathbf{r},t)$, where $\alpha_0$ determines the rate of release of chemicals (unit of $\alpha_0$: $[\text{length}]^{-2}$). From these equations one can derive the stochastic equation for the exact density following Dean's approach in Ref. \cite{Dean:1996}, which reads $\partial_t C =D \nabla^2 C + \mu \nabla \cdot (C \nabla \phi) + \nabla \cdot \left[\sqrt{2 D C} \; \mathbf{f}(\mathbf{r},t) \right]$, where $\mathbf{f}$ is a Gaussian white noise: $\langle f_\alpha(\mathbf{r}, t) f_\beta(\mathbf{r}', t')\rangle = \delta_{\alpha \beta} \delta(\mathbf{r}-\mathbf{r}') \delta(t-t')$. Note that this Langevin equation for the density is exact and contains the same information as the $N$-body stochastic Langevin equations \cite{Dean:1996, Chavanis:2010}. Moreover, using phenomenological arguments, one can use the same scheme for a continuous coarse-grained density $C(\mathbf{r},t)$ even when it cannot be described exactly as a sum of $\delta$-functions, while keeping track of the fluctuations \cite{Chavanis:2010}.

We now extend the model phenomenologically by adding a source term $L(C)$ that describes cell division and death and a noise $g(\mathbf{r},t)$ that breaks the cell number conservation:
\begin{eqnarray}
\partial_t C &=& D \nabla^2 C + \mu \nabla \cdot (C \nabla \phi) + \nabla \cdot \left[\sqrt{2 D C} \; \mathbf{f}(\mathbf{r},t) \right] \nonumber \\
&&+L(C)+\sqrt{2 M(C)} \;g(\mathbf{r},t),
\label{eq:c-eq}
\end{eqnarray}
where $\langle g(\mathbf{r}, t) g(\mathbf{r}', t')\rangle = \delta(\mathbf{r}-\mathbf{r}') \delta(t-t')$. We choose the logistic growth rule that corresponds to $L(C)=\lambda C(C_0-C)$ \cite{NowakBook}, where $C_0$ is the carrying capacity, and $\lambda$ is an effective growth rate per unit concentration (unit of $\lambda$: $[\text{length}]^d/[\text{time}]$). However, all our results are valid for any generic form for $L(C)$ provided $L(C_0) = 0$ and $\left. L'(C)\right|_{C_0} < 0$ \cite{Toner:2012}, since higher order terms in an expansion of the growth term renormalize to zero under RG. The strength of the nonequilibrium noise corresponding to number fluctuations, $M(C)$, is in general a function of the concentration and can be derived for any given form of the growth rule (unit of $M$: $[\text{time}]^{-1}\times[\text{length}]^{-d}$) \cite{Doering:2003}. For example, using a stochastic growth and coagulation process, one can derive the expression $M(C)=\lambda C (C_0+C)/2$ (see Appendix I).

To simplify the multiplicative noise term, we assume that the density fluctuates around a constant background of $C_0$; hence, we define the density via $C(\mathbf{r},t) = C_0 + \rho(\mathbf{r},t)$ and expand in $\rho/C_0$ up to the lowest order nonlinearity. Then the equation for the density fluctuations becomes
\begin{equation}
\partial_t \rho = D \nabla^2 \rho - \theta \rho - \nu_1 \nabla \cdot \left[\rho \nabla \left(\frac{1}{\nabla^2}\right) \rho \right] -\frac{\nu_2}{2}  \rho^2+ \eta,
\label{eq:rho-eq}
\end{equation}
where $\frac{1}{\nabla^2}$ is defined as the inverse Laplacian in Fourier space and the noise correlator in Fourier space is given as $\langle \eta(\mathbf{k}, \omega) \eta(\mathbf{k}', \omega') \rangle= 2 \left[D_0 + D_2 k^2\right] (2 \pi)^{d+1} \delta(\mathbf{k} + \mathbf{k}')\delta(\omega + \omega')$. The {\it bare} parameters are related to the physical characteristics of the system (introduced above) as follows: $\nu_1^{\rm bare}=\mu \alpha_0$, $\nu_2^{\rm bare}=2 \lambda$, $\theta^{\rm bare}=(\mu \alpha_0+\lambda) C_0$, $D_0^{\rm bare}=\lambda C_0^2$, and $D_2^{\rm bare}=D C_0$. Ignoring the nonlinearities, Eq. \eqref{eq:rho-eq} tells us that the uniform density phase is stable for $\theta > 0$, and signals a clumping instability that corresponds to chemotactic collapse at $\theta=0$, which corresponds to a new threshold of $\mu < \lambda/\alpha_0$ for chemotactic collapse for dividing cells, as opposed to $\mu=0$ \cite{KS1970}. The properties of the system at the phase transition could be studied by implementing a perturbative treatment of the nonlinear terms within a dynamical RG formulation \cite{Forster:1977,Medina:1989}.

\begin{figure}[t]
\centering \includegraphics[width = \columnwidth]{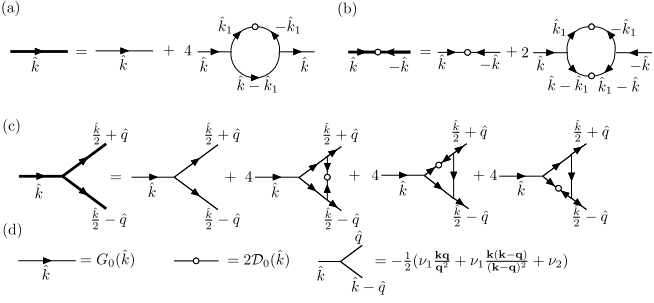}
\caption{One-loop diagrams (a) for the response function $G(\hat{k})$, (b) the noise correlator $\mathcal{D}(k)$, and (c) the Vertex function, in terms of the bare quantities defined in (d), namely $G_0(\hat{k})=[i \omega + D k^2 + \theta ]^{-1}$ and $\mathcal{D}_0(k) = D_0 + D_2 k^2$, where $\hat{k} := (\mathbf{k}, \omega)$.}
\label{fig:diagrams}
\end{figure}

The underlying assumption of RG calculations is invariance of Eq.~\eqref{eq:rho-eq} under rescaling space ($\mathbf{r}\rightarrow e^{\ell} \mathbf{r}$) and time ($\mathit{t}\rightarrow e^{\ell z} \mathit{t}$) close to a critical point. The cell density fluctuations will in this case transform as $\rho\rightarrow e^{\ell \chi} \rho$. The coarse-graining associated with the rescaling leads to corrections to the bare quantities in the Green function, noise correlator and the three-point function, which can be found by integrating out short-range degrees of freedom. The Feynman diagrams in Fig. \ref{fig:diagrams} provide a graphical representation of the lowest-order perturbative corrections to the bare quantities (see Appendix II). Coarse-graining corresponds to evaluating the (wavevector) integrals from $\Lambda e^{-\ell}$ to $\Lambda$ to eliminate large wavenumbers, where $\Lambda=2\pi/a$ is an upper cutoff in Fourier space and thus inversely proportional to a microscopic lengthscale $a$, which is set by the size of the cells.

We note that the cells will typically also experience short-range interactions, for example from excluded-volume effects. However, these interactions are irrelevant in RG sense as compared to the Coulomb-like interactions. To see this, consider adding to Eq. (\ref{eq:c-eq}) a term  of the form $\xi \nabla (C \nabla {\Psi})$ with ${\Psi} = \int d^d \mathbf{r}_0 C(\mathbf{r}_0,t) G(\mathbf{r} - \mathbf{r}_0))$ to account for short-range interactions ($\int G(\mathbf{r})d\mathbf{r} = {\rm const.}$). This term will scale as $b^{2 \chi-2}$, while the long-range interaction term scales as $b^{2 \chi}$. Therefore, we can ignore the short-range interactions for the purpose of studying the critical properties of the system.

Following the standard procedure briefly described above \cite{Forster:1977, Medina:1989}, we obtain the following RG flow equations for the coupling constants:

\begin{subequations}
\begin{widetext}
\begin{equation}
\begin{split}
\frac{d \theta}{d\ell} = z\theta - \frac{\nu_2 K_d \Lambda^{d-4}}{2 D^2}\Big\{ D_0 \Big[(3+2/d) \nu_1 + 2 \nu_2 \Big] + 3 \nu_1 D_2 \Lambda^2   \Big\},
\end{split}
\end{equation}
\vskip -0.5cm
\begin{equation}
\begin{split}
\frac{d \nu_1}{d\ell} = \nu_1\left\{\chi +  z + \frac{\nu_2 \pi K_d \Lambda^{d-6}}{4 D^3}\Big[3 \nu_1 + 2 \nu_2\Big] (D_0 + D_2 \Lambda^2)  \right\},
\end{split}
\end{equation}
\vskip -0.5cm
\begin{equation}
\begin{split}
\frac{d \nu_2}{d\ell} = \nu_2\left\{\chi +  z + \frac{\pi K_d \Lambda^{d-6}}{ D^3}\Big[3 \nu_1 + 2 \nu_2\Big] \Big[\nu_1 + \nu_2\Big] (D_0 + D_2 \Lambda^2)  \right\},
\end{split}
\end{equation}
\vskip -0.5cm
\begin{equation}
\begin{split}
\frac{d D}{d\ell} = D \left(z - 2 - \frac{K_d \Lambda^{d-6}}{8 D^3}\Bigg\{ \frac{(2 d-4)}{d} D_2 \Lambda^2 \Big[3 \nu_1 - 2 \nu_2\Big] \Big[2 \nu_1 + \nu_2 \Big] + 4 D_0 \nu_2 \Big[\frac{(17 - 5 d)}{d} \nu_1 + 2 \nu_2\Big]\Bigg\}\right),
\end{split}
\end{equation}
\vskip -0.5cm
\begin{equation}
\begin{split}
&\frac{d D_0}{d\ell} = D_0 (z - d - 2 \chi) + \frac{\nu_2^2 K_d \Lambda^{d-6}}{2 D^3} (D_0 + D_2 \Lambda^2)^2,
\end{split}
\end{equation}
\vskip -0.5cm
\begin{equation}
\begin{split}
\frac{d D_2}{d\ell} =  D_2 (z - d - 2 - 2 \chi)  - \frac{\nu_2 K_d \Lambda^{d-8}}{8 d D^3}\Big\{  &7 \nu_1\Big[D_0^2 + \frac{(8 d -2)}{7} D_0 D_2 \Lambda^2 + D_2^2 \Lambda^4\Big] \\&+ \nu_2 (D_0 + D_2 \Lambda^2) \Big[(3d-14) D_0 + (d-2) D_2 \Lambda^2 \Big] \Big\},
\end{split}
\end{equation}
\end{widetext}
\label{eq:flowequations}
\end{subequations}
where $K_d={S_d}/{(2\pi)^d}$ and $S_d=2 \pi^{d/2}/\Gamma(d/2)$ is the area of unit sphere in $d$ dimensions.

\begin{figure}[t]
\includegraphics[width = 0.85 \columnwidth]{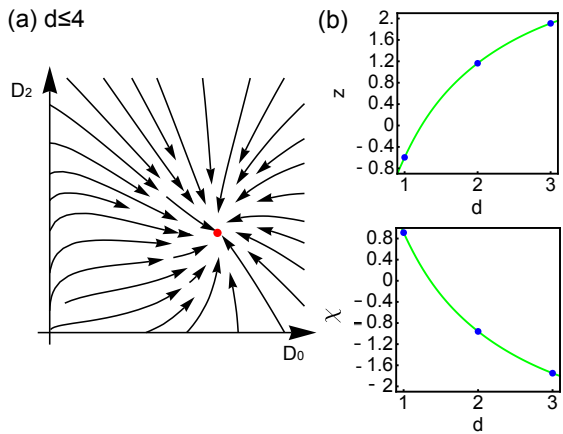}
\caption{(a) The fixed-point is stable in the space of equilibrium and nonequilibrium noises for $d \leq 4$. (b) The values for the exponents $z$ and $\chi$ corresponding to the stable fixed point. For $d=(1,2,3)$ dimensions, the values are reported in Table \ref{tab:param}.}
\label{fig:summary22}
\end{figure}

Scale invariance at the critical point requires that the values of the coupling constants in Eq.~\eqref{eq:rho-eq} remain fixed under RG flow.
This condition gives a nontrivial fixed point [see Fig.~\ref{fig:summary2}(a)] corresponding to fixed point values
$\nu_1^* = -3 \sqrt{D^3 (\chi +z)/\left[\pi K_d (D_0+D_2 \Lambda^2)\right]}$ and $\nu_2^* = -\frac{4}{3} \nu_1^*$, indicating that both chemotaxis and cell growth are relevant for the collective dynamics of cells at the collapse transition. Around the fixed point, we find a flow depicted in Fig.~\ref{fig:summary2}(a). Starting with $\nu_1$ and $\nu_2$ below a dimension-dependent threshold value indicated by a separatrix in Fig.~\ref{fig:summary2}(a), the flow will go towards $\nu_1^*$ and $\nu_2^*$. Above the threshold, however, the system will be controlled by a perturbatively non-accessible fixed point with runaway behaviour. Runaway flows have been commonly associated with first-order transitions \cite{Iacobson:1981} but this is not guaranteed unless the flow goes to a region where a first order transition can be established using a mean-field approximation \cite{Cardy:1995}. Here, the flow will move the system to a region where noise and growth dominate and chemotaxis is irrelevant. Thus, above the threshold, Eq.~\eqref{eq:rho-eq} will belong to the class of Fisher equations, which are known to exhibit instabilities, nonlinear fronts and chaos \cite{Kessler:1998, Conlon:2005, Hallatschek:2009}. To decide on the nature of the phase transition, an analysis of the Fisher-type equation under consideration of short-range interactions is needed. Below the threshold we also find that the fixed point value of the tuning parameter is renormalized as $\theta^*= -\frac{2D}{\pi}\Lambda^2\left(\frac{\chi + z}{z}\right)\left[1+\frac{6}{d} \frac{D_0}{D_0 + D_2 \Lambda^2}\right]$. Hence, the system will be controlled by the nontrivial fixed point if the bare values of the nonlinear terms correspond to the basin of attraction of the fixed point and we tune $\theta$ to its fixed point value. Combining these conditions with the dependence of the bare coupling constants on the known microscopic parameters of the system (see above), yields the phase diagram that is depicted in Fig.~\ref{fig:summary2}(b) for different dimensions $d$. The flow equations for the noise strengths $D_0$ and $D_2$ lead to a (perturbatively accessible) nontrivial stable fixed point in the physically relevant dimensions ($d\leq 4$) as shown in Fig. \ref{fig:summary22}(a). The fixed point has both equilibrium and nonequilibrium components, highlighting that even a starting point with a bare value for noise that has no nonequilibrium component, i.e. $G=0$, this type of noise will be generated through the RG process.

Figure~\ref{fig:summary22}(b) shows the values for the exponents $z$ and $\chi$, which are numerically exact within this order of perturbation theory, and the values for dimensions $d=(1,2,3)$ are tabulated in Table \ref{tab:param}. These critical exponents imply anomalous diffusion and density fluctuations, which could be experimentally probed. The single cell mean-square displacement behaves as $\Delta L^2(t) = \langle [\mathbf{r}(t) - \mathbf{r}(0)]^2\rangle\sim t^{2/z} = t^{\alpha}$, which implies superdiffusive behaviour in two and three dimensions (see Table \ref{tab:param}). The scaling form of density correlations $\langle \rho(\mathbf{r},t)\rho(\mathbf{r}',t) \rangle \sim |\mathbf{r} - \mathbf{r}'|^{2 \chi}$ can be used to calculate the overall number fluctuations $\Delta N^2=\int d^d \mathbf{r} d^d \mathbf{r}' \langle \rho(\mathbf{r},t)\rho(\mathbf{r}',t) \rangle $ leading to $\Delta N \sim L^{d+\chi}=L^{\beta}$. Note that for a system of particles with no interaction and cell division the exponents $\chi = -d$ and $z=2$ would hold, which correspond to normal diffusion and no scaling for number fluctuations. The calculations yield a negative exponent $z$ in a one-dimensional system, which would result in finite mean-square displacements corresponding to localization of cells.

\begin{table}
\caption{This table summarizes the critical exponents found at the critical fixed point.}
\begin{ruledtabular}
\begin{tabular}{cccc}
 & $d=1$ & $d=2$ & $d=3$ \\
\hline
$z$ & $-0.60$ & $1.16$ & $1.90$ \\
$\chi$ & $0.91$ & $-0.95$ & $-1.75$ \\
$\alpha = 2/z$ & $-3.33$ & $1.72$ & $1.05$ \\
$\beta = d+\chi$ & $1.91$ & $1.05$ & $1.25$
\label{tab:param}
\end{tabular}
\end{ruledtabular}
\end{table}

Our predictions could be tested in experiments if it is possible to control the parameters involved such as the cell carrying capacity $C_0$, the growth rate $\lambda$ or the diffusion constant $D$. From Fig.~\ref{fig:summary2}(b) one can see that the phase transition could be detected when the average time a cell takes to move by a distance of the order of its own size is comparable with the rate of cell division. For tumor cells, typical cell division rates are of the order of $10^{-5}/s$ \cite{Pearlman:1976}. With an estimated viscosity of soft tissues of $1 \; {\rm Pa}\; {\rm s}$ \cite{Eskandari:2008} and a cell body length of the order of $10 \; \mu {\rm m}$, the estimated diffusion constant will be $\sim 10^{-4} \; \mu {\rm m}^2/{\rm s}$, which shows that the desired order of magnitude for the effect is well within reach. We could speculate that the separatrix in Fig.~\ref{fig:summary2}(a) might be interpreted as signaling a transition to a phase where growth and number fluctuations dominate. Characterization of this transition, however, is beyond the scope of our simple perturbative description. Interestingly, the chemotactic interactions are prominent in controlling the large-scale behaviour of the system together with the cell division dynamics.

In summary, we have applied dynamical RG to study the collective behaviour of cells which undergo chemotaxis and division, and are under the influence of non-number-conserving nonequilibrium noise. We find a rich phase diagram, which in some parameter regime predicts a critical phase transition with nontrivial exponents that can be calculated perturbatively, and suggest anomalous diffusion of cells and long-range correlations. Our results might help shed light on the question of what controls the communication between strongly dividing cells that are far apart and their collective behaviour. This could help us towards addressing the fundamentally challenging questions of what determines the sharp onset of metastasis and how the metacommunity of metastatic cells across the body of a cancer patient could still coordinate their activities \cite{Hanahan:2011}.

This work is supported by Ernst Ludwig Ehrlich Studienwerk (AG), EPSRC (AG), and Human Frontier Science Program (HFSP) grant RGP0061/2013 (RG). RG thanks the KITP for hospitality and acknowledges funding by NSF grant PHY11-25915.

\vspace{0.5cm}
\noindent
\textbf{APPENDIX I: Derivation of the Nonequilibrium Noise}

\vspace{0.3cm}

As discussed, the exact form of the density dependence of the noise that describes number fluctuations will depend on the specific process.
Here we use a simple model to study the interplay between stochastic growth and coagulation that leads to death, and use it to derive a noisy logistic 
growth equation \cite{Doering:2003}. We will then use the microscopic calculation to derive an expression for the noise strength $M$. 

Let us first consider a volume $V$ in which we study the growth-coagulation process described as
\begin{subequations}
\begin{equation}
A  \rightarrow A + A \quad \text{with rate }\Gamma,
\label{growthrate2}
\end{equation}
\begin{equation}
A  + A \rightarrow A \quad \text{with rate }\Upsilon,
\end{equation}
\end{subequations}
where $\Upsilon/V$ is taken to be the coagulation rate of distinct pairs of particles within the volume. The stochastic equation for the probability $P(n, t)$ for having $n$ particles at time $t$ in that volume reads
\begin{equation}
\begin{split}
&\partial_t P(n, t) =\\
& \Gamma (n-1) P(n-1, t) + \frac{1}{2} \left(\frac{\Upsilon}{V}\right) (n+1) n P(n+1, t)\\
& - \Gamma n P(n, t) + \frac{1}{2} \left(\frac{\Upsilon}{V}\right) n (n-1) P(n, t).
\end{split}
\end{equation}
This can the be expanded up to second order in $n$, which results in the Fokker-Planck equation
\begin{equation}
\begin{split}
&\partial_t P(n, t)=-\partial_n \left[\Gamma n P(n, t) - \frac{\Upsilon}{2} n \left(\frac{n-1}{V}\right)\right] P(n, t) \\
&+ \frac{1}{2}\partial_n^2 \left[\Gamma n P(n, t) + \frac{\Upsilon}{2} n \left(\frac{n-1}{V}\right)\right] P(n, t).
\end{split}
\end{equation}
This will correspond to the Langevin equation for the density $C = n/V$ as
\begin{equation}
\begin{split}
&\frac{dC(t)}{dt} = \left(\Gamma+\frac{\Upsilon}{2V}\right) C(t) -\frac{\Upsilon}{2} C(t)^2 \\
&+ \frac{1}{\sqrt{V}} \sqrt{\left(\Gamma - \frac{\Upsilon}{2 V}\right) C(t) + \frac{\Upsilon}{2} C(t)^2}\; \zeta(t),
\end{split}
\end{equation}
where $\langle \zeta(t) \zeta(t') \rangle = \delta(t-t')$. 

We can now generalize the scheme to take into account spatial structure, by treating this result as corresponding to the concentration in the $i$th volume element, namely
\begin{equation}
\begin{split}
&\frac{dC_i(t)}{dt} = \left(\Gamma+\frac{\Upsilon}{2V}\right) C_i(t) -\frac{\Upsilon}{2} C_i(t)^2\\
&+ \frac{1}{\sqrt{V}} \sqrt{\left(\Gamma - \frac{\Upsilon}{2 V}\right) C_i(t) + \frac{\Upsilon}{2} C_i(t)^2}\; \zeta_i(t),
\end{split}
\end{equation}
where $\langle \zeta_i(t) \zeta_j(t') \rangle = \delta_{ij} \delta(t-t')$. Taking the continuum limit, we will have $C_i(t) \to C({\bf r},t)$, $\zeta_i(t) \to \zeta({\bf r},t)$, and $\delta_{ij} \to V \delta(\mathbf{r}-\mathbf{r}')$, which altogether yields
\begin{equation}
\begin{split}
&\frac{\partial C({\bf r},t)}{\partial t} = \left(\Gamma+\frac{\Upsilon}{2V}\right) C({\bf r},t) -\frac{\Upsilon}{2} C({\bf r},t)^2\\
&+ \sqrt{\left(\Gamma - \frac{\Upsilon}{2 V}\right) C({\bf r},t) + \frac{\Upsilon}{2} C({\bf r},t)^2}\; \zeta({\bf r},t),
\end{split}
\end{equation}
$\langle \zeta(\mathbf{r}, t) \zeta(\mathbf{r}', t')\rangle = \delta(\mathbf{r}-\mathbf{r}') \delta(t-t')$. This is of the form
\begin{equation}
\begin{split}
\frac{\partial C({\bf r},t)}{\partial t} = \lambda C(C_0-C) + \sqrt{\lambda \left(C_0-\frac{2}{V}\right) C + \lambda C^2} \; \zeta(\mathbf{r}, t),
\end{split}
\end{equation}
where $\lambda =\frac{1}{2}\Upsilon$ and $C_0 = \frac{2 \Gamma}{\Upsilon}+ \frac{1}{V}$. Assuming that at saturation there are many more cells in the volume element than 2 (i.e. $C_0 \gg 2/V$), we obtain $M(C)=\lambda C (C_0+C)/2$ as reported in the main text.

\vspace{0.3 cm}
\noindent
\textbf{APPENDIX II: Details of the Diagrammatic Expansion}

\vspace{0.3 cm}

In this section, we explain how we perform the perturbative expansion that is sketched in Fig. 3 Fourier transforming Eq.~(2) gives
\begin{widetext}
\begin{equation}
\begin{split}
  i \omega \rho(\mathbf{k}, \omega ) =  - D k^2 \rho( \hat{k}) - \theta \rho( \hat{k}) + \eta(\hat{k}) -\frac{1}{2} \int  \frac{d^{d+1}\hat{k}_1}{ (2 \pi)^{d+1}}  \rho(\hat{k}_1 )\rho( \hat{k} - \hat{k}_1) \times \left[\nu_1\mathbf{k}\cdot(\mathbf{k} - \mathbf{k}_1) /(\mathbf{k} - \mathbf{k}_1)^2 + \nu_1\mathbf{k} \cdot \mathbf{k}_1 /\mathbf{k}_1^2 + \nu_2\right]
\label{eq:rho-ft}
\end{split}
\end{equation}
We used the convention $\rho(\mathbf{r},t) = \int d \hat{k} /(2 \pi)^{d+1} \rho(\hat{k}) e^{i \omega t - i \mathbf{k} \mathbf{x}}$. For brevity, $\hat{k}$ is defined as $\hat{k} := (\mathbf{k}, \omega)$. One can define the bare Green's function $G_0(\hat{k})=[i \omega + D k^2 + \theta ]^{-1}$, which will diverge for $\theta = 0$ at large length- and long time-scales ($\omega \rightarrow 0 \text{, } \mathbf{k}\rightarrow 0$). In additon, we define the bare noise correlator $\mathcal{D}_0(k) = D_0 + D_2 k^2$ and the effective vertex function $\Gamma_0(\mathbf{k}, \mathbf{q}) = \Gamma_0(\mathbf{k}, \mathbf{k}-\mathbf{q}) = -\frac{1}{2}(\nu_1 \mathbf{k} \cdot \mathbf{q} /(\mathbf{q})^2 + \nu_1 \mathbf{k} \cdot(\mathbf{k} - \mathbf{q})(\mathbf{k} - \mathbf{q})^2 + \nu_2)$.

This divergence indicates that the large-scale behaviour of the cells will be self-similar for $\theta \rightarrow 0$. In the following, we will focus solely on this case. We rewrite Eq.~\eqref{eq:rho-ft} as
\begin{equation}
\begin{split}
\rho(\hat{k}) = G_0(\hat{k}) \eta(\hat{k}) - \frac{1}{2} G_0(\hat{k}) \int \frac{d^{d+1}\hat{k}_1}{(2 \pi)^{d+1}} \rho(\hat{k}_1)\rho( \hat{k} - \hat{k}_1) \times \big[ \nu_1 \mathbf{k}(\mathbf{k} - \mathbf{k}_1) /(\mathbf{k} - \mathbf{k}_1)^2 +  \nu_1 \mathbf{k}\mathbf{k}_1 /(\mathbf{k}_1)^2 + \nu_2\big]
\label{eq:rho-ft2}
\end{split}
\end{equation}
Equation \eqref{eq:rho-ft2} is a convenient starting point for a diagrammatic expansion in orders of $\nu_{1,2}$. From this we find the effective response function $G(\hat{k})$ (defined by $\rho(\hat{k})=G(\hat{k}) \eta(\hat{k})$), the effective noise correlator $\mathcal{D}(k)$ and the effective vertex function $\Gamma$. The corresponding one-loop Feynman diagrams, which are shown in Fig.~3, translate to the following explicit expressions:
\begin{subequations}
\small
\begin{equation}
\begin{split}
&G(\hat{k})^{-1} =  G_0(\hat{k})^{-1} - \int \frac{d^{d+1} \hat{k}_1}{(2 \pi)^{d+1}}\Big\{  \mathcal{D}(\mathbf{k}_1)\big[  \nu_1 \mathbf{k} \cdot (\mathbf{k} - \mathbf{k}_1)/(\mathbf{k} - \mathbf{k}_1)^2 + \nu_1 \mathbf{k} \cdot \mathbf{k}_1 /(\mathbf{k}_1)^2 + \nu_2\big] \\
&\times\big[ \nu_1(\mathbf{k}-\mathbf{k}_1) \cdot \mathbf{k} /(\mathbf{k})^2 - \nu_1(\mathbf{k}-\mathbf{k}_1)\cdot \mathbf{k}_1 /(\mathbf{k}_1)^2 + \nu_2 \big] |G_0(\hat{k}_1)|^2 G_0(\hat{k}-\hat{k}_1)                         \\
&+ \mathcal{D}(\mathbf{k}-\mathbf{k}_1)  \big[\nu_1 \mathbf{k}  \cdot   (\mathbf{k} - \mathbf{k}_1)/(\mathbf{k} - \mathbf{k}_1)^2 + \nu_1 \mathbf{k} \cdot \mathbf{k}_1 /(\mathbf{k}_1)^2 + \nu_2\big]\\
&\times\big[ \nu_1 \mathbf{k}_1 \cdot \mathbf{k} A(\mathbf{k}) - \nu_1\mathbf{k}_1 \cdot (\mathbf{k} - \mathbf{k}_1) /(\mathbf{k}-\mathbf{k}_1)^2 + \nu_2 \big] |G_0(\hat{k}-\hat{k}_1)|^2 G_0(\hat{k}_1) \Big\}
\end{split}
\label{eq:g}
\end{equation}
\begin{equation}
\begin{split}
&2 \mathcal{D}(\mathbf{k})  = 2 \mathcal{D}_0(\mathbf{k}) + \frac{1}{2} \int \frac{d^{d+1} \hat{k}_1}{(2 \pi)^{d+1}} \Big\{ 2 \mathcal{D}_0 (\mathbf{k}_1) 2 \mathcal{D}_0(\mathbf{k}-\mathbf{k}_1)\big[ \nu_1\mathbf{k}\cdot (\mathbf{k} - \mathbf{k}_1)/(\mathbf{k} - \mathbf{k}_1)^2 + \nu_1 \mathbf{k} \cdot \mathbf{k}_1 /(\mathbf{k}_1)^2 + \nu_2 \big] \\
&\times \left[ \nu_1 \mathbf{k}\cdot(\mathbf{k}-\mathbf{k}_1)/(\mathbf{k}-\mathbf{k}_1)^2 + \nu_1 \mathbf{k} \cdot \mathbf{k}_1 /(\mathbf{k}_1)^2 + \nu_2 \right] |G_0(\hat{k}_1)|^2 |G_0(\hat{k}-\hat{k}_1)|^2  \Big\}
\end{split}
\end{equation}
\label{eq:correlators}
\begin{equation}
\begin{split}
&\Gamma(\mathbf{k}, \mathbf{k}/2 + \mathbf{q})  = \Gamma_0(\mathbf{k}, \mathbf{k}/2 + \mathbf{q}) -  \int \frac{d^{d+1} \hat{q}_1}{(2 \pi)^{d+1}}\\ 
&\Big\{ \mathcal{D} (\mathbf{q} - \mathbf{q}_1) 
\big[ \nu_1 \frac{\mathbf{k}\cdot (\mathbf{k}/2 + \mathbf{q}_1)}{(\mathbf{k}/2 + \mathbf{q}_1)^2} + \nu_1 \frac{\mathbf{k} \cdot (\mathbf{k}/2 - \mathbf{q}_1)}{(\mathbf{k}/2 - \mathbf{q}_1)^2} + \nu_2 \big]\times  \big[ \nu_1\frac{ (\mathbf{k}/2 + \mathbf{q}_1)\cdot (\mathbf{k}/2 + \mathbf{q})}{(\mathbf{k}/2 + \mathbf{q})^2} + \nu_1 \frac{(\mathbf{k}/2 + \mathbf{q}_1) \cdot (\mathbf{q}_1 - \mathbf{q})}{(\mathbf{q}_1 - \mathbf{q})^2} + \nu_2 \big]\times \\
&      \big[ \nu_1\frac{(\mathbf{k}/2 - \mathbf{q}_1)\cdot (\mathbf{k}/2 - \mathbf{q})}{(\mathbf{k}/2 - \mathbf{q})^2} + \nu_1 |\frac{(\mathbf{k}/2 - \mathbf{q}_1) \cdot (\mathbf{q} - \mathbf{q}_1)}{(\mathbf{q} - \mathbf{q}_1)^2} + \nu_2 \big]   \times G_0(\hat{k}/2 + \hat{q}_1) G_0(\hat{k}/2 - \hat{q}_1) |G_0(\hat{q}-\hat{q}_1)|^2 
\\
&+ 2\times  \mathcal{D} (\mathbf{k}/2 + \mathbf{q}_1) 
\big[ \nu_1\frac{ \mathbf{k}\cdot (\mathbf{k}_2 + \mathbf{q}_1)}{(\mathbf{k}/2 + \mathbf{q}_1)^2} + \nu_1 \frac{\mathbf{k} \cdot (\mathbf{k}/2 - \mathbf{q}_1)}{(\mathbf{k}/2 - \mathbf{q}_1)^2} + \nu_2 \big] \times          \big[ \nu_1\frac{(\mathbf{q} - \mathbf{q}_1)\cdot (\mathbf{k}/2 + \mathbf{q})}{(\mathbf{k}/2 + \mathbf{q})^2} + \nu_1 \frac{(\mathbf{q} - \mathbf{q}_1) \cdot (\mathbf-{k}/2 - \mathbf{q}_1)}{(\mathbf{k}/2 + \mathbf{q}_1)^2} + \nu_2 \big] \times     \\
& \big[ \nu_1\frac{(\mathbf{k}/2 - \mathbf{q}_1)\cdot (\mathbf{k}/2 - \mathbf{q})}{(\mathbf{k}/2 - \mathbf{q})^2 }+ \nu_1 \frac{(\mathbf{k}/2 - \mathbf{q}_1) \cdot (\mathbf{q} - \mathbf{q}_1)}{(\mathbf{q} - \mathbf{q}_1)^2} + \nu_2 \big]\times G_0(\hat{k}/2 + \hat{q}_1) |G_0(\hat{k}/2 - \hat{q}_1) |^2 G_0(\hat{q}-\hat{q}_1)\Big\}
\end{split}
\end{equation}
\label{eq:vertex}
\end{subequations}
\end{widetext}
\normalsize
The rescaling of length- and timescales in general leads to corrections of the bare quantities in Eq.~\eqref{eq:correlators}. These corrections can be found by integrating out short-range degrees of freedom. If the scaling factor $b$ is chosen as $b = e^{\ell}$ this corresponds to evaluating the integrals in Eqs.~\eqref{eq:correlators} from $\Lambda e^{-\ell}$ to $\Lambda$, to eliminate large wavelengths \cite{Forster:1977}.

\end{document}